\documentclass[12pt]{iopart}

\usepackage{graphicx}
\usepackage{color}

\begin{document}

\title[]{Non-equilibrium vibrational and electron energy distributions functions in atmospheric nitrogen ns pulsed discharges and $\mu$s post-discharges: the role of electron molecule vibrational excitation scaling-laws}

\author{Gianpiero Colonna$^1$, Vincenzo Laporta$^{1,2}$, Roberto Celiberto$^{3,1}$, Mario Capitelli$^{4,1}$ and Jonathan Tennyson$^2$}
\address{$^1$ Istituto di Metodologie Inorganiche e dei Plasmi, CNR, Via Amendola 122/D, 70125 Bari, Italy}
\address{$^2$ Department of Physics and Astronomy, University College London, London WC1E 6BT, UK}
\address{$^3$ Dipartimento di Ingegneria Civile, Ambientale, del Territorio, Edile e di Chimica, Politecnico di Bari, 70125 Bari, Italy}
\address{$^4$ Dipartimento di Chimica, Universit\`{a} di Bari, Via Orabona, 4, 70126, Bari Italy}
\ead{gianpiero.colonna@imip.cnr.it}
\vspace{10pt}
\begin{indented}
\item[]October 2014
\end{indented}

\begin{abstract}
The formation of the electron energy distribution function in nanosecond atmospheric nitrogen discharges is investigated by means of self-consistent solution of the chemical kinetics and the Boltzmann equation for free electrons. The post-discharge phase is followed to few microseconds. The model is formulated in order to investigate the role of the cross section set, focusing on the vibrational-excitation by electron-impact through resonant channel. Four different cross section sets are considered, one based on internally consistent vibrational-excitation calculations which extend to the whole vibrational ladder, and the others obtained by applying commonly used scaling-laws.
\end{abstract}

%
%
%
%
%

\section{Introduction\label{sec:intro}}

A lot of work has been presented over the last 40 years aimed at understanding the non-equilibrium vibrational kinetics of non-equilibrium molecular plasmas~\cite{0963-0252-11-3A-302, 0963-0252-16-1-S03, 0029-5515-46-6-S06} due to their applications in many technological fields including plasma-assisted combustion~\cite{Starikovskiy201361, 0022-3727-39-16-R01}, negative ion sources~\cite{Hassouni1998502, 1556673, 0029-5515-46-6-S05, :/content/aip/journal/pop/17/6/10.1063/1.3431635}, excimer lasers~\cite{106836}, microelectronics~\cite{refId0}, aerospace~\cite{doi:10.1021/jp048847v, Armenise1996} and environment applications~\cite{Colonna2001587}.

In particular, nitrogen containing plasmas have been extensively considered as N$_2$ is the main component of air. Particular attention has been devoted to the coupling of vibrational and electronic distributions with the electron energy distribution function through the so-called superelastic collisions (also known as second kind collisions)~\cite{Cacciatore1982141, Gorse198863}. The treatment of non-equilibrium vibrational and electronic kinetics has become progressively more complex, with many elementary processes included to better model the plasma~\cite{0022-3727-23-11-005, Colonna19935, JAP:8075438, :/content/aip/journal/pop/21/12/10.1063/1.4904817, 0963-0252-20-2-024007, 0963-0252-1-3-011, Laux201246, 0963-0252-18-3-034023}. Sophisticated models are now available which can be used to understand the properties of nitrogen containing plasmas. Unfortunately the available experimental measurements generally characterize average quantities, such as the degree of dissociation and ionization, rather than the more informative vibrational distribution functions (vdf) and electron energy distribution functions (eedf), which remain difficult to measure. The recent measurements of the vdf of nitrogen~\cite{0022-3727-32-15-317, 0022-3727-34-12-307} are limited to vibrational levels with $v<20$ and thus fail to probe the plateau predicted for the higher lying vibrational levels. Similarly, the scant experimental information on eedf emphasizes the post-discharge regime. Different sets of cross sections as well as different forms of non-equilibrium vibrational distributions were used by Dyatko \emph{et al.}~\cite{0022-3727-26-3-011, 1221831} to reproduce experimental vibrational temperatures in low pressure nitrogen afterglows. More recently Dyatko \emph{et al.} have developed a complete plasma kinetic model determining the cross sections for electronic excitation for $v>0$ shifting the known cross sections for the ground vibrational level and making use of Franck Condon factors~\cite{Dyatko:2010aa}. A number of theoretical calculations of the vdf and eedf are available which are generally characterized by the use of different data sets for the key cross sections. In this context interesting results were reported for the low pressure nitrogen afterglow by Guerra \emph{et al.} by coupling eedf and electronically excited state kinetics to reproduce the experimental eedf from second derivatives of digitized probe characteristics measured using a triple probe technique~\cite{1221827}.

The accuracy and the completeness of the existing data bases is not well-known and hard to estimate. A sensitivity analysis would be welcome to elucidate the key processes which determine the form of both the vdf and eedf. One process is of paramount importance for coupling the two distributions, namely the electron-vibration (e-V) processes \emph{i.e.} the pumping of vibrational quanta over the vibrational ladder through resonant electron-molecule interaction. For this process a complete set of cross sections interconnecting the whole N$_2$ vibrational ladder through single- and multi-quantum transitions has recently become available~\cite{0963-0252-21-5-055018, 0963-0252-23-6-065002}. These e-V cross sections are indeed essential for determining the initial distribution of vibrational quanta over the vibrational ladder and also, at sufficiently large electron densities, for giving the spread of these quanta over the whole vibrational ladder. In the high electron density case, the e-V process plays the same role of the vibration-vibration (V-V) energy transfer processes.

In the past, the lack of a complete cross section database forced researchers to either limit their treatment of e-V processes to those interconnecting the first 8-10 vibrational levels \cite{PhysRevA.13.188, 0022-3727-32-20-307} or to use scaling-laws of unknown accuracy to interconnect levels with $v> 10$. Recently, a cross section data set due to Huo \emph{et al.}~\cite{Huo_rate} was combined with extended experimental results to derive a new scaling-law~\cite{ISI:000090142100005}. It should be noted that a similar scaling-law was proposed by Gordiets \emph{et al.} limiting however its application to $v=10$~\cite{467998}. However, all previous models suffer from the incompleteness of the relevant e-V cross section set; this can now be addressed by use of the recent, comprehensive compilation of Laporta \emph{et al.}~\cite{0963-0252-21-5-055018, 0963-0252-23-6-065002}.

The e-V processes compete with the V-V and vibration-translation (V-T) energy transfer processes in the formation of the non-equilibrium vdf of nitrogen, where the V-T process includes as deactivating partners both molecules (V-Tm) and atoms (V-Ta). V-V and V-Tm rates have been and are being continuously updated by different groups. The fitted rates based on the pioneering work by Billing \emph{et al.}~\cite{raey, Capitelli1980299} are still considered as benchmark values for the numerous other calculations appearing in the literature. We note that the first studies on the non-equilibrium vibrational kinetics including e-V over the first few levels, V-V and V-Tm over the whole vibrational ladder yielded the well-known Treanor-Gordiets distribution \emph{i.e.} a vibrational distribution containing essentially a Treanor law over the first few vibrational levels, followed by a long plateau for higher $v$ which finally ends in a Boltzmann distribution at the gas temperature. The interplay between V-V and V-Tm rates determined the end of the plateau and the onset of a vdf at the gas temperature $T_g$. Under favourable conditions the long plateau speed-up the dissociation process as well as the ionization one, as a result of heavy particle collisions between vibrationally excited molecules (the so-called `pure vibrational mechanisms'). In these models, V-Ta were constrained to have the same rates as V-Tm \emph{i.e.} were inefficient at destroying the vibrational content of nitrogen molecules. This assumption persisted until Lagana \emph{et al.}~\cite{doi:10.1021/j100286a015, doi:10.1021/j100053a025} presented quasi-classical calculations of the V-Ta processes based on a semi-empirical potential energy surface (PES). These data were first used by Armenise \emph{et al.}~\cite{Armenise1992597} for describing the non-equilibrium vdf of nitrogen under discharge and aerospace re-entry conditions. The use of the new rates for discharge conditions yielded to a decrease of the length of the plateau in the vibrational distribution due to the interplay of V-V and V-Ta processes. The pure vibrational mechanisms were seen to lose their importance in the dissociation and ionization processes. A problem however arises when using the V-Ta rates of Lagana \emph{et al.} at low temperature, because these rates were limited to $v\leq 10$. This problem has been resolved by the extensive quasi-classical trajectory (QCT) calculations on the same process by Esposito \emph{et al.}~\cite{Esposito20061, Esposito199949, Esposito2000193} who used the same PES as Lagana \emph{et al.} These new data, while basically confirming the V-Ta rates of Lagana \emph{et al.} for $v > 10$, are orders of magnitude lower than the extrapolated Lagana \emph{et al.} rates for $v < 10$. These differences are extremely important in determining the behaviour of low-lying vibrationally excited levels of N$_2$ which has consequences for the kinetics of the whole system. The use of the new e-V and V-Ta cross sections and rates is therefore important to illuminate the role of the corresponding processes in affecting the non-equilibrium vibrational kinetics of nitrogen plasmas. In particular, they are likely to key for determining structures in both the vdf and eedf, with consequences for the dissociation and ionization rates.

To avoid possible compensation effects with other important elementary processes, we base our study on a kinetic code with all other input data fixed at that already adopted by our group. Furthermore, to avoid the need for full dimensional fluid dynamics, we use a zero-dimensional code based on the coupling of vdf, eedf and electronically excited state kinetics. In particular, we restrict our analysis to the role of e-V processes in affecting vdf and eedf of nitrogen discharges \emph{i.e.} using for the VTa the recent analytical curves of Esposito \emph{et al.}~\cite{Esposito20061}.

The aim of this paper is to apply these ideas to nanosecond discharges followed by a post-discharge analysis of about 1$\ \mu$s. In particular, we focus our attention on the formation of the eedf under discharge and post-discharge conditions, and the creation of structures in the eedf due to superelastic collisions from vibrationally and electronically excited states. Particular emphasis is given to the dependence of computed microscopic and macroscopic quantities on the electron-molecule resonant vibrational-excitation cross sections. The paper is divided in three sections: The Section \ref{sec:th_eN2_xsec} deals with a comparison of the complete e-V data base with approximate scaling laws given in the literature. The Section \ref{sec:model} is computation of the eedf and vdf in the nanosecond and microsecond regimes, and its dependence on the e-V cross sections. Conclusions and perspectives are reported in the Section \ref{sec:conclusions}.

\section{Theoretical electron-nitrogen cross sections\label{sec:th_eN2_xsec}}

Laporta \emph{et al.}~\cite{0963-0252-21-5-055018, 0963-0252-23-6-065002} calculated low-energy resonant cross sections and corresponding rate constants for electron-N$_2$. Vibrational-excitation transitions between all vibrational levels of ground state N$_2$, parameterized on rotational quantum number $J$, with $\Delta J = 0$, were considered, using the following scheme:
\begin{equation}
e + \textrm{N}_2(\textrm{X}\,^1\Sigma^+_g,v,J) \to \textrm{N}^-_2(^2\Pi_g) \to e + \textrm{N}_2(\textrm{X}\,^1\Sigma^+_g,v',J)\,, \label{eq:eN2_RVE}
\end{equation}
where the $\textrm{N}^-_2(^2\Pi_g)$ is the short-lived, anionic resonance state. In this section some results are reviewed and, in particular, the linear scaling-law between rate constants are critically discussed in the light of this new data.

\begin{figure}
\begin{indented}
\item[]
\begin{tabular}{cc}
\includegraphics[scale=.7,angle=0]{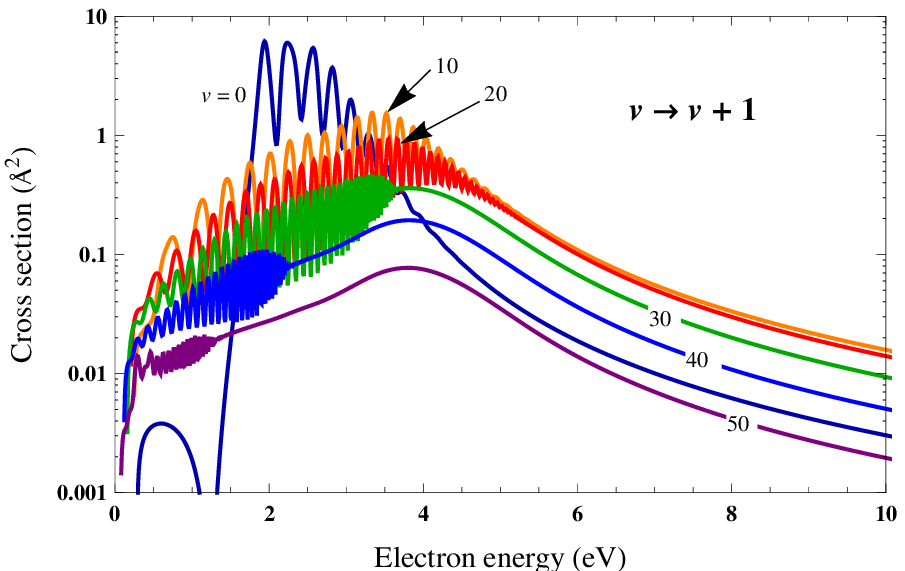} & \includegraphics[scale=.7,angle=0]{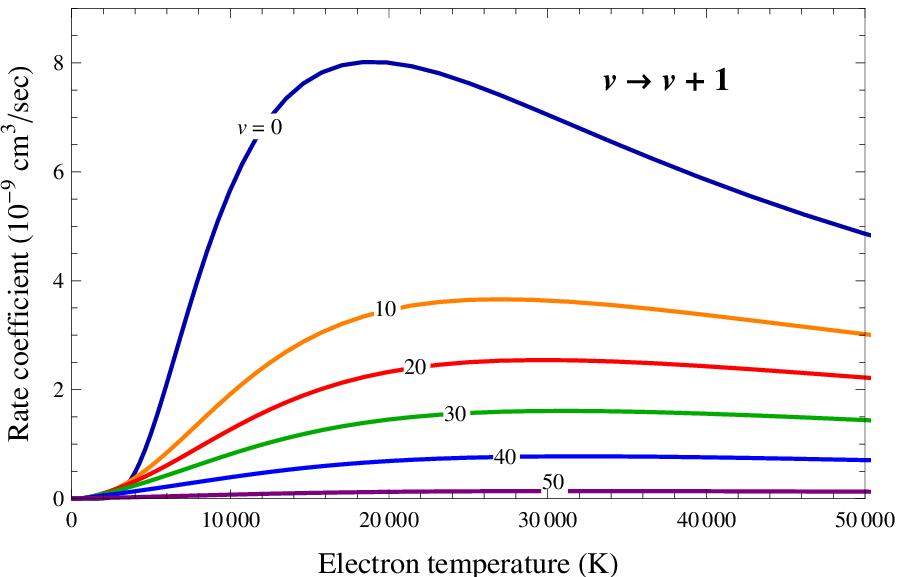}\\
\includegraphics[scale=.7,angle=0]{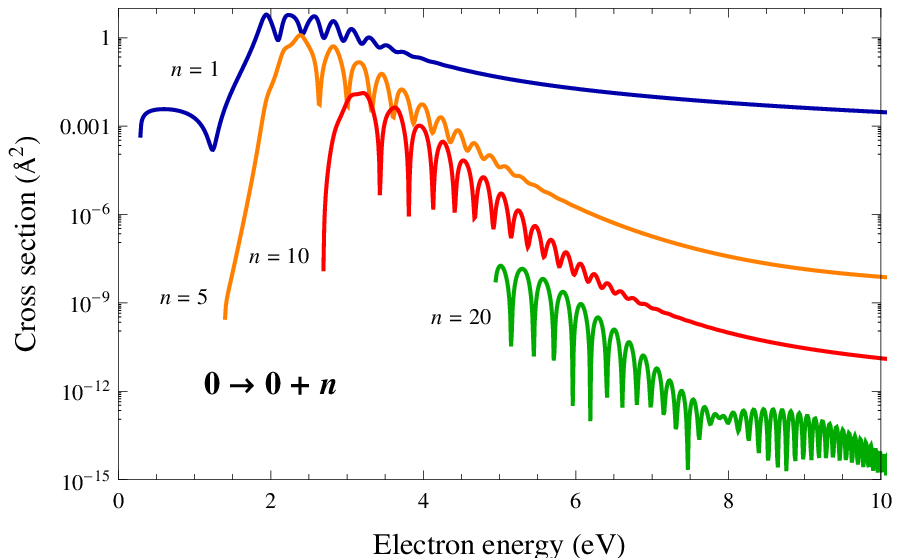} & \includegraphics[scale=.7,angle=0]{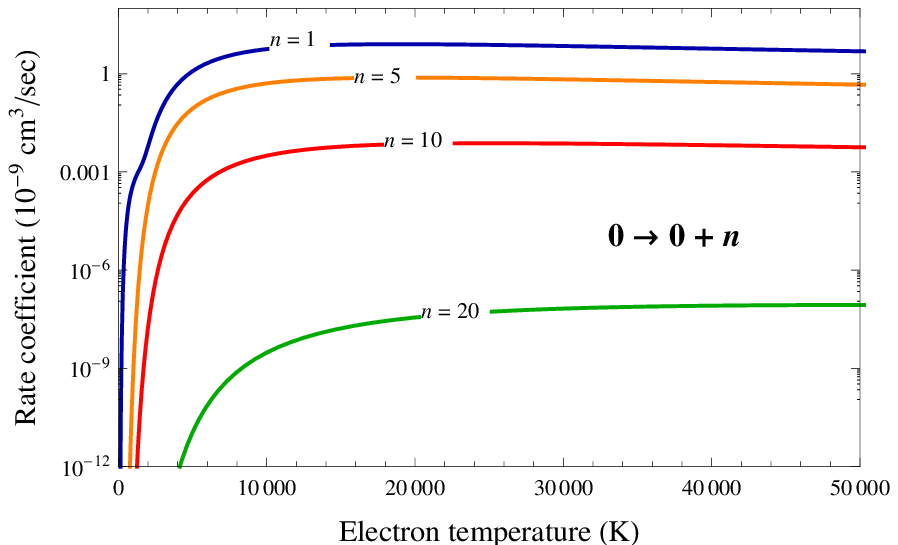}\\
\includegraphics[scale=.7,angle=0]{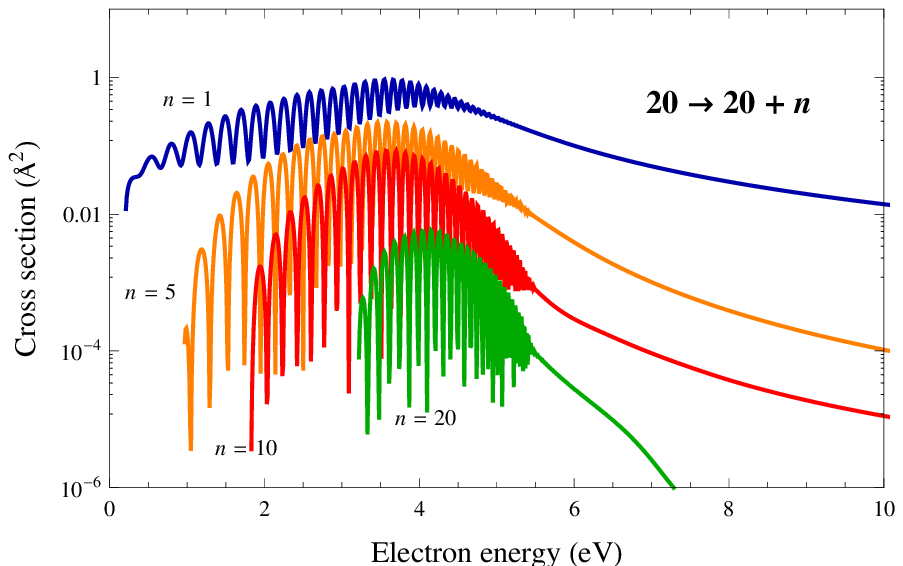} & \includegraphics[scale=.7,angle=0]{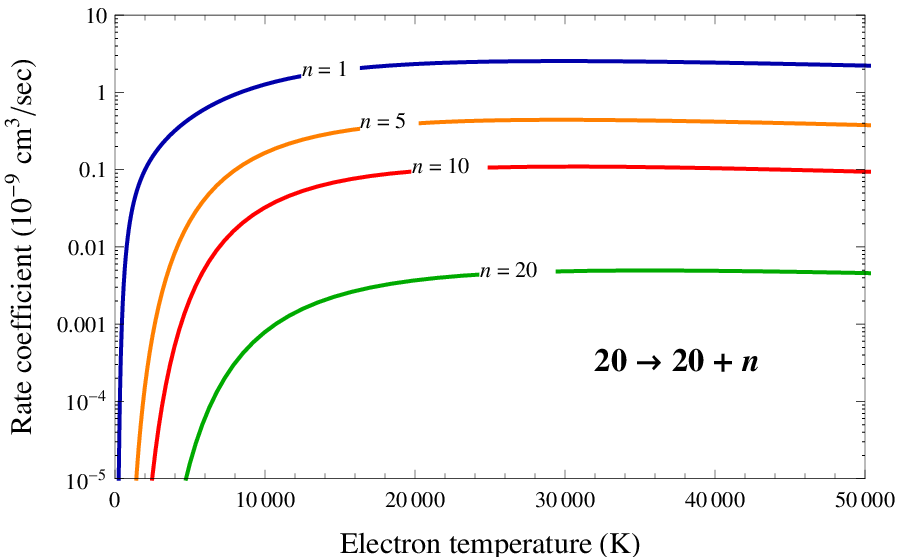}
\end{tabular}
\end{indented}
\caption{Some series of cross sections (on the left) and the corresponding rate coefficients (on the right) for resonant vibrational-excitation of N$_2$ by electron-impact calculated by Laporta \emph{et al.}~\cite{0963-0252-21-5-055018, 0963-0252-23-6-065002}. \label{fig:N2xsec}}
\end{figure}

Laporta \emph{et al.} used a phenomenological potential energy curve for the N$_2$ electronic ground state which supports 68 vibrational levels for the $J = 0$ rotation state. The cross section calculations were performed using a projection operator formalism within the local approximation~\cite{wadehra} to solve the nuclear dynamics. Figure \ref{fig:N2xsec} shows some results of cross sections for single-quantum transitions (upper frame) and for $n$-quantum transition from the $v = 0$ (middle frame) and from the $v = 20$ states (lower frame). Typical resonant cross sections are characterized by a series of narrow peaks that correspond to the vibrational levels of the resonant state. The corresponding rate constants, see right-side of Fig.~\ref{fig:N2xsec}, are calculated as a function of the electron temperature by the convolution of the relevant cross section with a Maxwell distribution for the electrons. The complete set of cross sections and rate constants are available from Phys4Entry database~\cite{F4Edatabase}.

\begin{figure}
\begin{indented}
\item[]
\begin{tabular}{cc}
\includegraphics[scale=.7,angle=0]{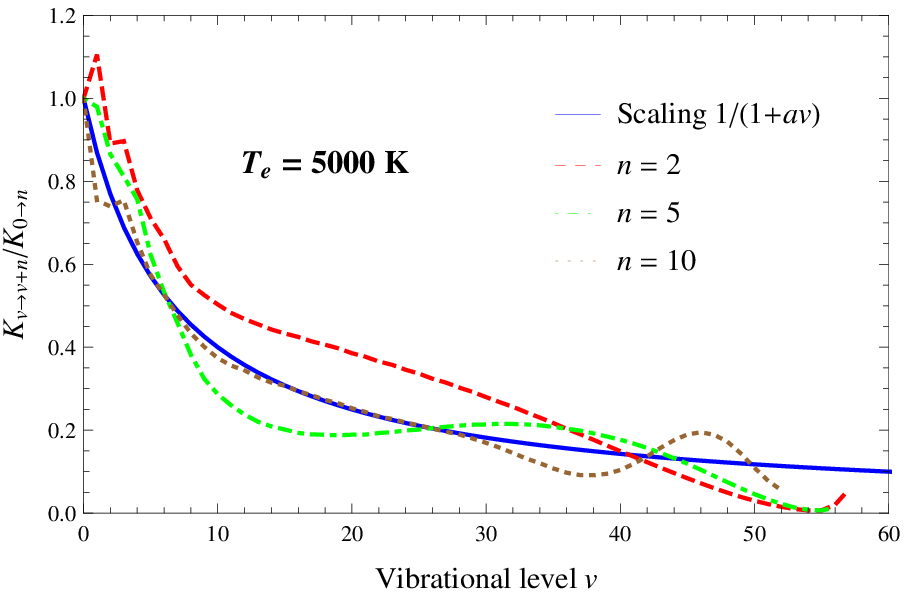}  & \includegraphics[scale=.7,angle=0]{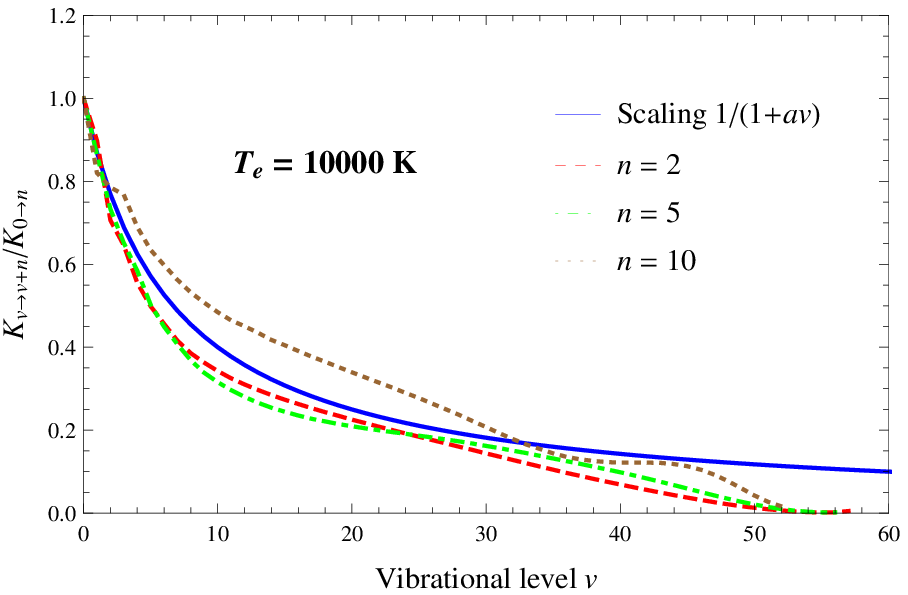}
\end{tabular}
\end{indented}
\caption{Comparison between the ratio $K_{v,v+n}/K_{0,n}$ and the scaling-law in Eq.~(\ref{eq:scalinglaw}) for $T_e = 5000$~K and $T_e = 10000$~K, for different vibrational $n$-quantum jumps, as a function of the initial vibrational state. The solid line is the scaling factor $1/(1+av)$. \label{fig:scalinglaw}}
\end{figure}

As already noted, a complete, internally-consistent, data set of cross sections for electron impact vibrational-excitation was not previously available. To complete the available data different scaling-laws have been applied. The first scaling-law we consider was reported in Ref.~\cite{ISI:000090142100005} and was obtained by fitting the data calculated by Huo \emph{et al.}~\cite{Huo_rate}. The resulting expression gives rate coefficients for transitions involving excited vibrational levels, starting from the rate for the $v = 0$ level:
\begin{equation}
K_{v,v+n} = \frac{K_{0,n}}{1+av} \,,\label{eq:scalinglaw}
\end{equation}
where $a = 0.15$. Figure \ref{fig:scalinglaw} shows the ratio $K_{v,v+n} / K_{0,n}$ as a function of the vibrational level $v$ for different $n$-quantum jumps at a fixed electron temperature $T_e$. We note that the scaling relation of Eq.~(\ref{eq:scalinglaw}) is a good approximation at high electron temperature, but loses accuracy for high vibrational states.

Another simple scaling-law, reported in Ref.~\cite{JAP:8075438}, is given by:
\begin{equation}
\sigma_{v,v+n}(\epsilon) = \sigma_{0,n}(\epsilon-\Delta\epsilon)\,,
\label{eq:scalinglaw_sigma}
\end{equation}
where $\Delta\epsilon$ is the energy difference due to the anharmonicity of the vibrational ladder:
\begin{equation}
\Delta\epsilon = \epsilon_n -\epsilon_0 + \epsilon_{n+v} - \epsilon_v\,.
\end{equation}
Only transitions with $v+n < 10$ have been considered. A similar approach has been used in Refs.~\cite{Cacciatore1982141, :/content/aip/journal/jcp/82/4/10.1063/1.448374}, where $\Delta\epsilon$ were neglected.

A third approach was obtained by extending the scaling-law for V-Tm rates for harmonic oscillator to the electron impact rates \emph{i.e.} it was applied a Laundau-Teller model to the e-V rates:
\begin{equation}
K_{v,v+1} = (v+1)K_{0,1}\,.\label{eq:scalinglaw3}
\end{equation}
In the next section we compare results obtained from four different kinetic models which use, respectively, the scaling laws in Eqs.~(\ref{eq:scalinglaw})-(\ref{eq:scalinglaw3}) given above and the full set of calculated cross sections by Laporta \emph{et al.}

\section{Kinetic model results \label{sec:model}}

The kinetic model used in this paper has been described in detail elsewhere~\cite{Colonna1996, Colonna2001, :/content/aip/journal/pop/20/10/10.1063/1.4824003, 0022-3727-34-12-308, Capitelli201431}. Briefly it consists of homogeneous kinetic equations coupling the non-equilibrium vibrational and electronically-excited state kinetics with the Boltzmann equation for the eedf. The processes induced by heavy particle collisions are listed in Table~1 of Ref.~\cite{:/content/aip/journal/pop/20/10/10.1063/1.4824003}. In particular, the processes coupling the vdf with electronically excited states and the chemical processes (dissociation and ionization) involving electronically excited states should be noted.  The list of electron-impact transitions with vibrationally excited N$_2$ molecules is reported in Table~2 of Ref.~\cite{:/content/aip/journal/pop/20/10/10.1063/1.4824003}.

The coupling between the chemical kinetics and the Boltzmann equation for free electrons is self-consistent, \emph{i.e.} at each time step the level population and the gas composition give the input quantities for the Boltzmann equation and the rate coefficients for the electron induced processes are calculated from the actual eedf. Inelastic and superelastic collisions involving vibrationally and electronically excited states and free electrons are fully considered in this model. Also electron-electron collisions are included using an efficient algorithm described in Ref.~\cite{Angola20101204}. The N$_2$ vibrational ladder includes 68 levels. This requires a rescaling of the old V-V and V-Tm rates which are based on 46 vibrational levels.

To solve the coupled problem we impose initial conditions for all the relevant quantities. In particular a Boltzmann distribution function at $T = 500$~K was selected for vdf and a Maxwell distribution function at the same temperature was selected for eedf. An initial molar fraction of electrons of $\chi_e = 10^{-10}$ was selected while pressure and gas temperature were kept at $p = 1$~bar and $T_g = 500$~K. Moreover, an uniform reduced electric field of $E/N = 200$~Td was applied for 3~ns and the simulation was followed in the post-discharge ($E/N=0$) for up to few microseconds \emph{i.e.} we are using an Heavyside-tipe temporal evolution for $E/N$. The simulation is therefore typical for a nanosecond high voltage-high pressure discharge followed by a post-discharge regime typically met in plasma-assisted combustion. This kind of simulation can be justified in the framework of the aim of the present paper mainly devoted to the understanding of the role of different scaling laws in the e-V cross sections. More realistic $E/N$ forms derived from the experiments~\cite{0022-3727-39-16-R01, 0022-3727-47-11-115201,  0741-3335-57-1-014001} including a rise time of the applied voltage followed by a plateau as well as more complicated forms are left to future work.

We consider four test cases, see Table~\ref{tab:test_cases}, based on the use of different scaling-laws for generating the electron-collision vibrational-excitation cross sections and on the numerically calculated data of Laporta \emph{et al}. Cross sections for all the other process, including the heavy particle kinetics, are the same for all the four cases. It should be noted that only models A and D include all the transitions between vibrational states, while model C considers transitions for levels with $v \leq 10$ and single-quantum excitations only.
\begin{table}
\begin{indented}
\item[]
\begin{tabular}{cl}
\hline
Case & Set of cross sections considered\\
\hline \hline
A & Scaling-law of Eq. (\ref{eq:scalinglaw}) is applied to the cross sections \cite{ISI:000090142100005}\\
B & Scaling-law of Eq. (\ref{eq:scalinglaw_sigma}) is used to calculate cross sections up to $v = 10$ \cite{JAP:8075438}\\
C & Scaling-law of Eq. (\ref{eq:scalinglaw3}) is applied to the cross sections\\
D & Full set of cross sections taken from Ref.~\cite{0963-0252-21-5-055018, 0963-0252-23-6-065002}\\
\hline
\end{tabular}
\end{indented}
\caption{Scheme of the models used in the calculations. \label{tab:test_cases}}
\end{table}

Before examining the results we estimate the characteristic relaxation times for the eedf and vdf. The eedf, neglecting vibrationally excited molecules, reaches a quasi-stationary state in a timescale of the order,
\begin{equation}
t_{qse} = (NK_{eV})^{-1} \approx 0.3\ \textrm{ns} \,,\label{eq:tqse}
\end{equation}
where $N = 1.52\times 10^{19}$~cm$^{-3}$ is the number density of heavy particles at $p = 1$~bar and $T = 500$~K and $K_{eV}\approx 2\times 10^{-8}$~cm$^{-3}$ is the rate coefficient for pumping vibrational energy in the system at $E/N = 200$~Td including all transitions from $v = 0$.

\begin{figure}
\begin{indented}
\item[]
\begin{tabular}{cc}
\includegraphics[scale=.35,angle=0]{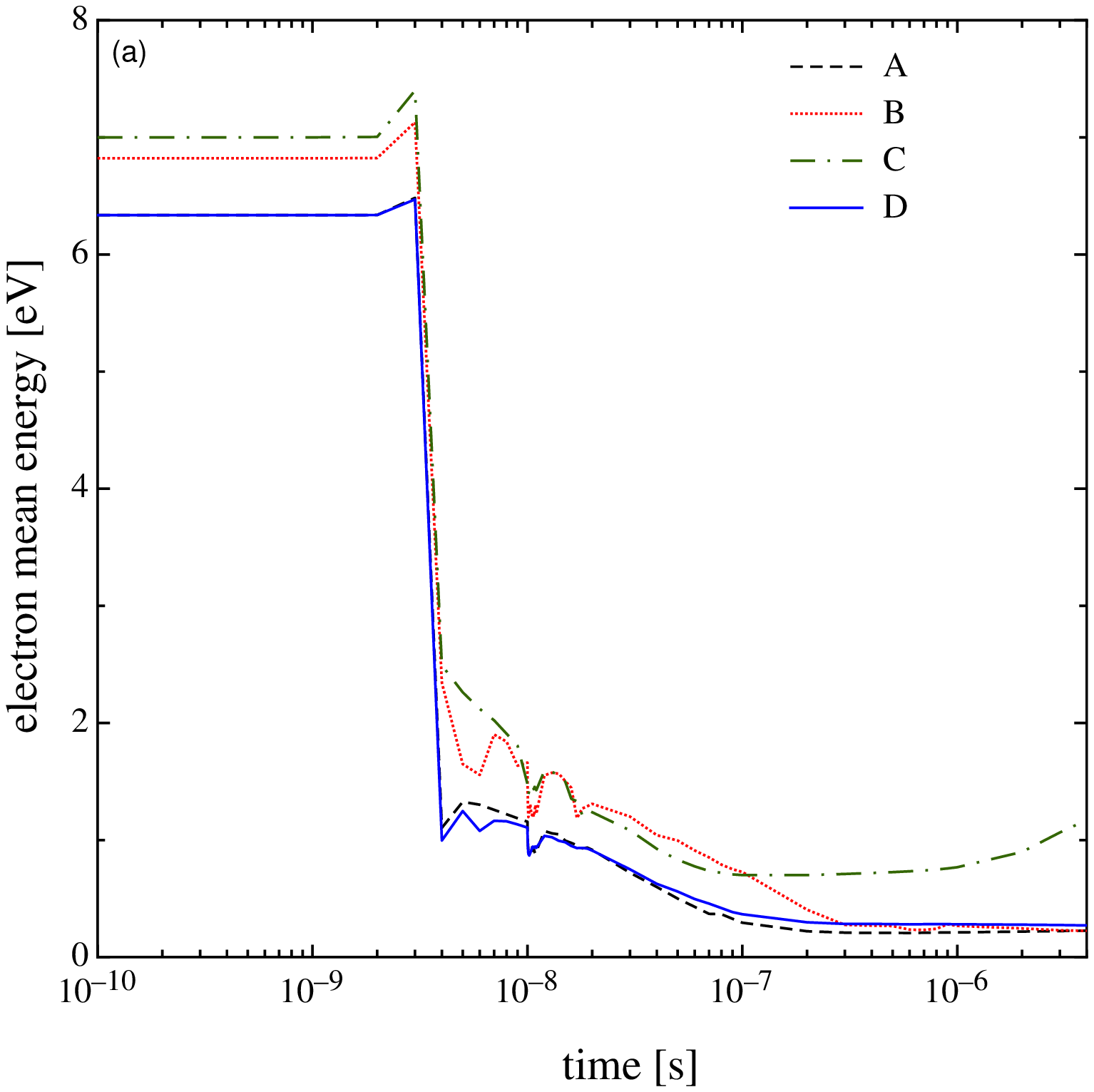} & \includegraphics[scale=.35,angle=0]{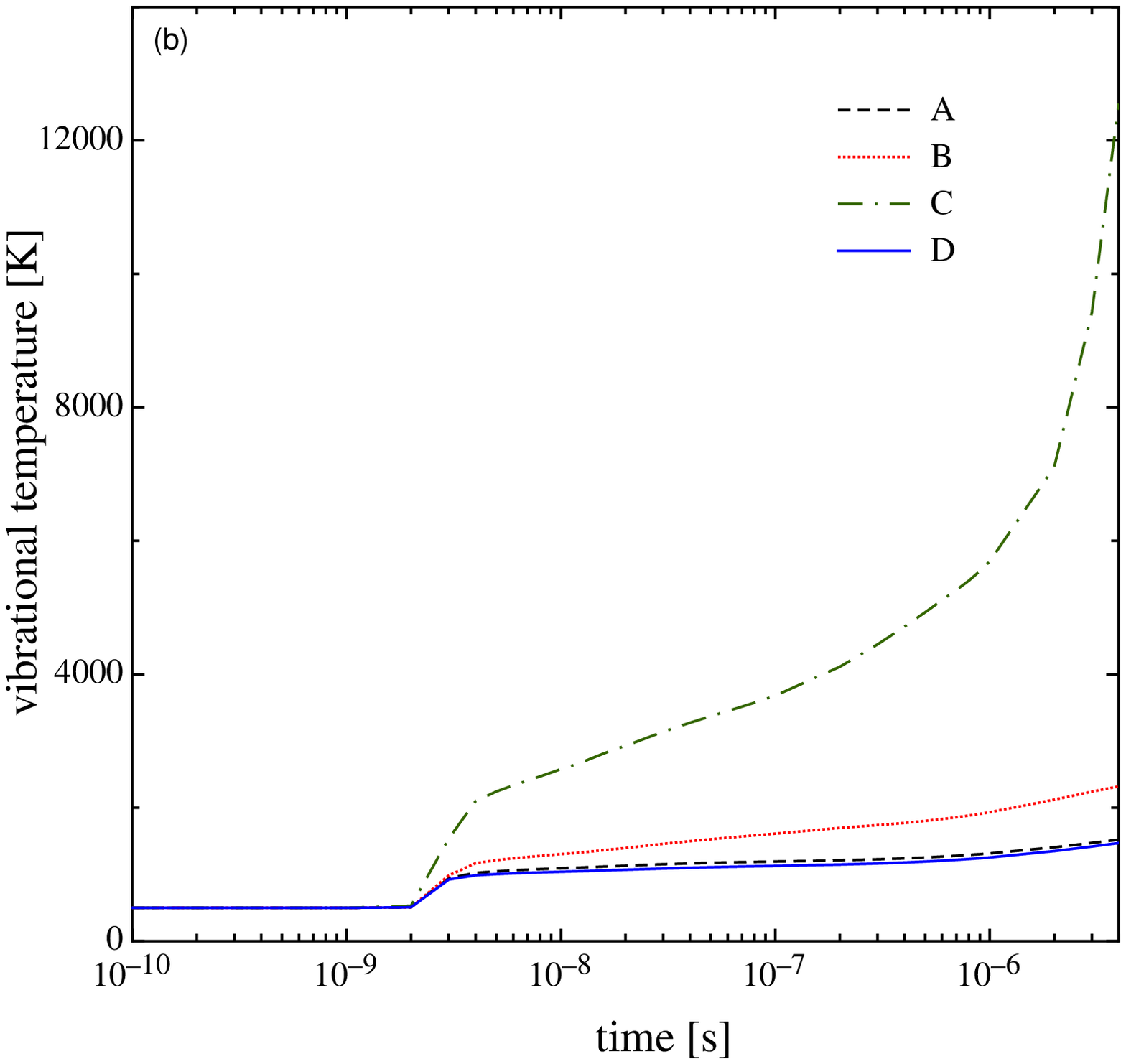} \\
\includegraphics[scale=.35,angle=0]{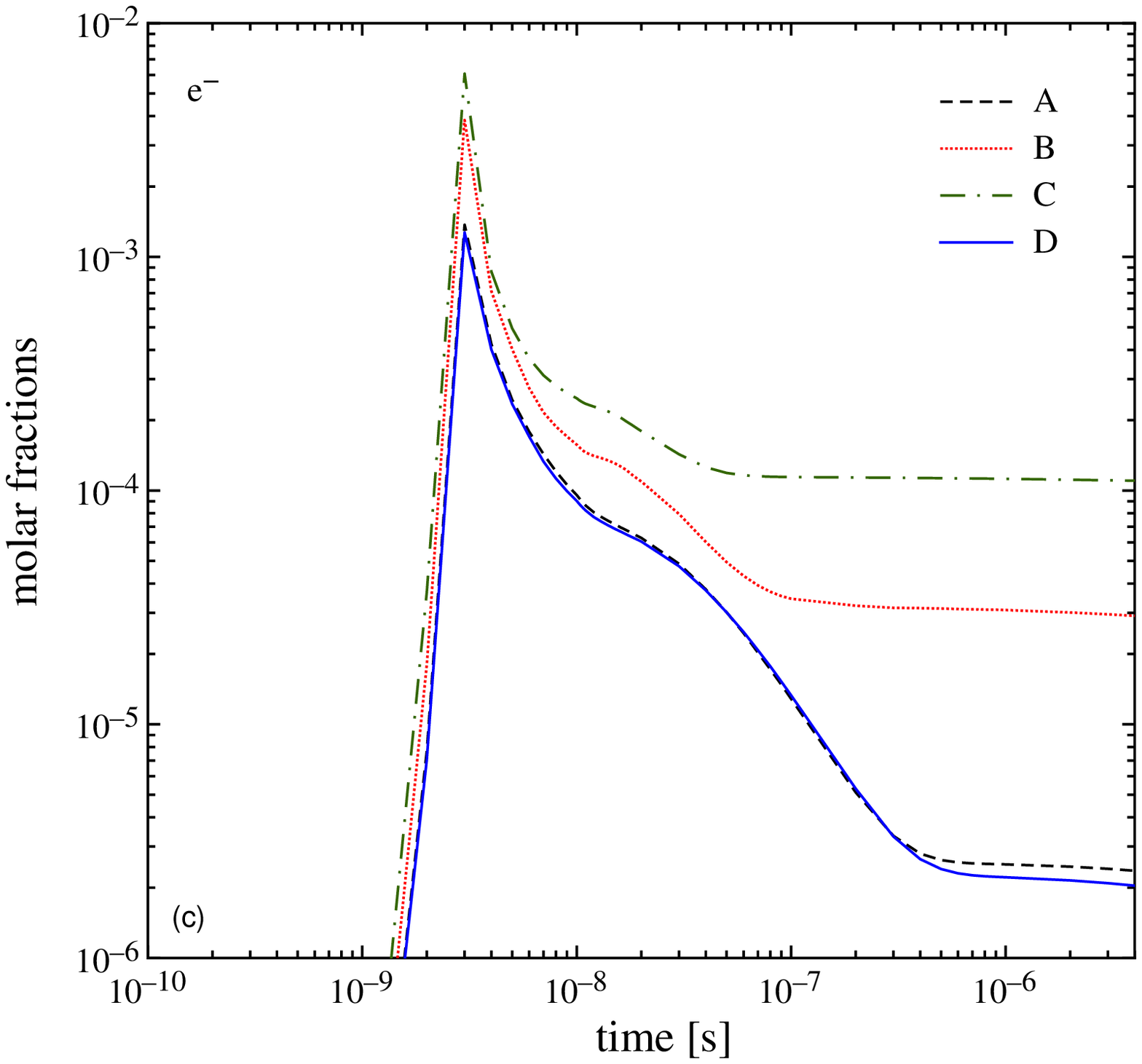} & \includegraphics[scale=.35,angle=0]{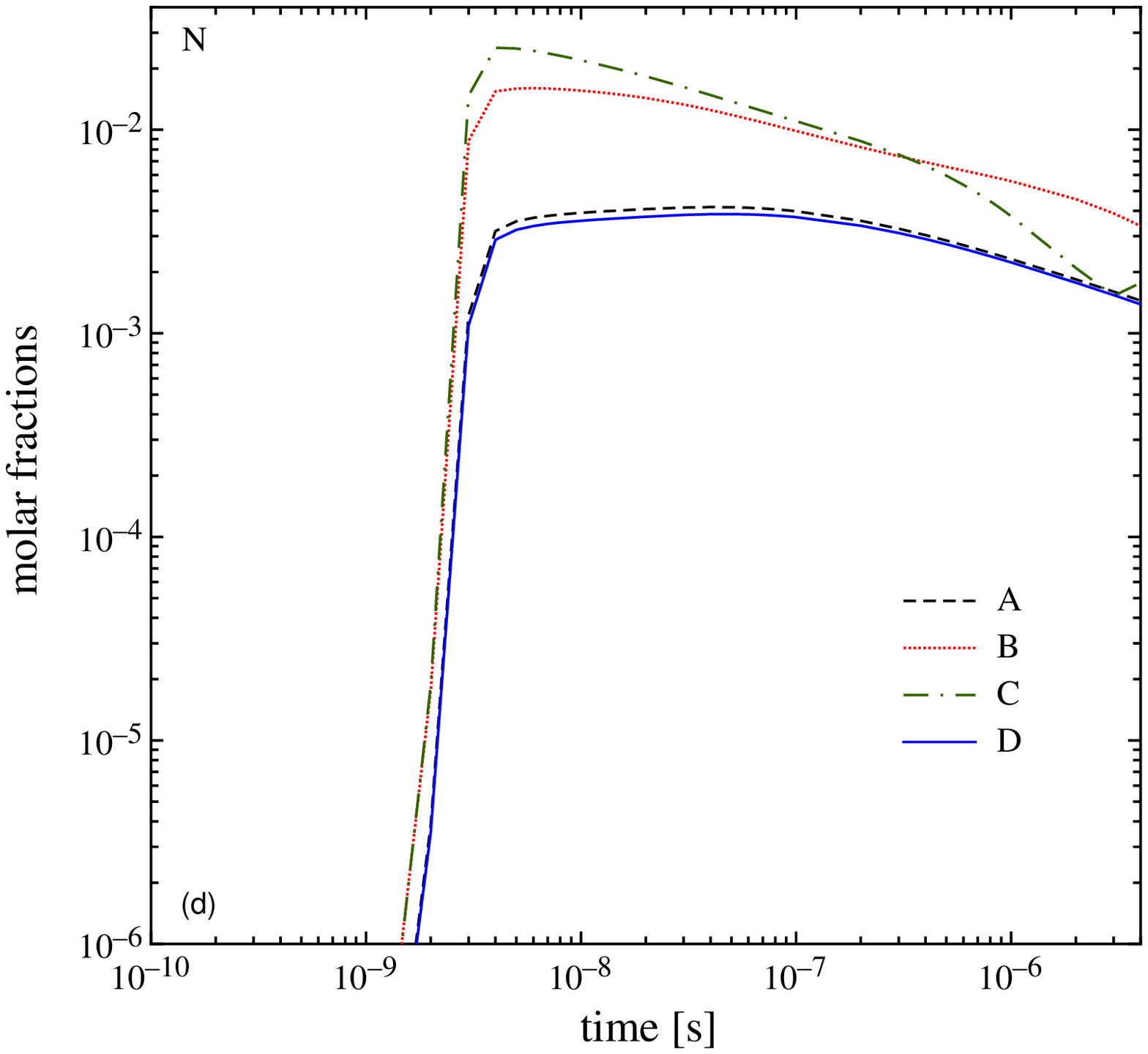} \\
\includegraphics[scale=.35,angle=0]{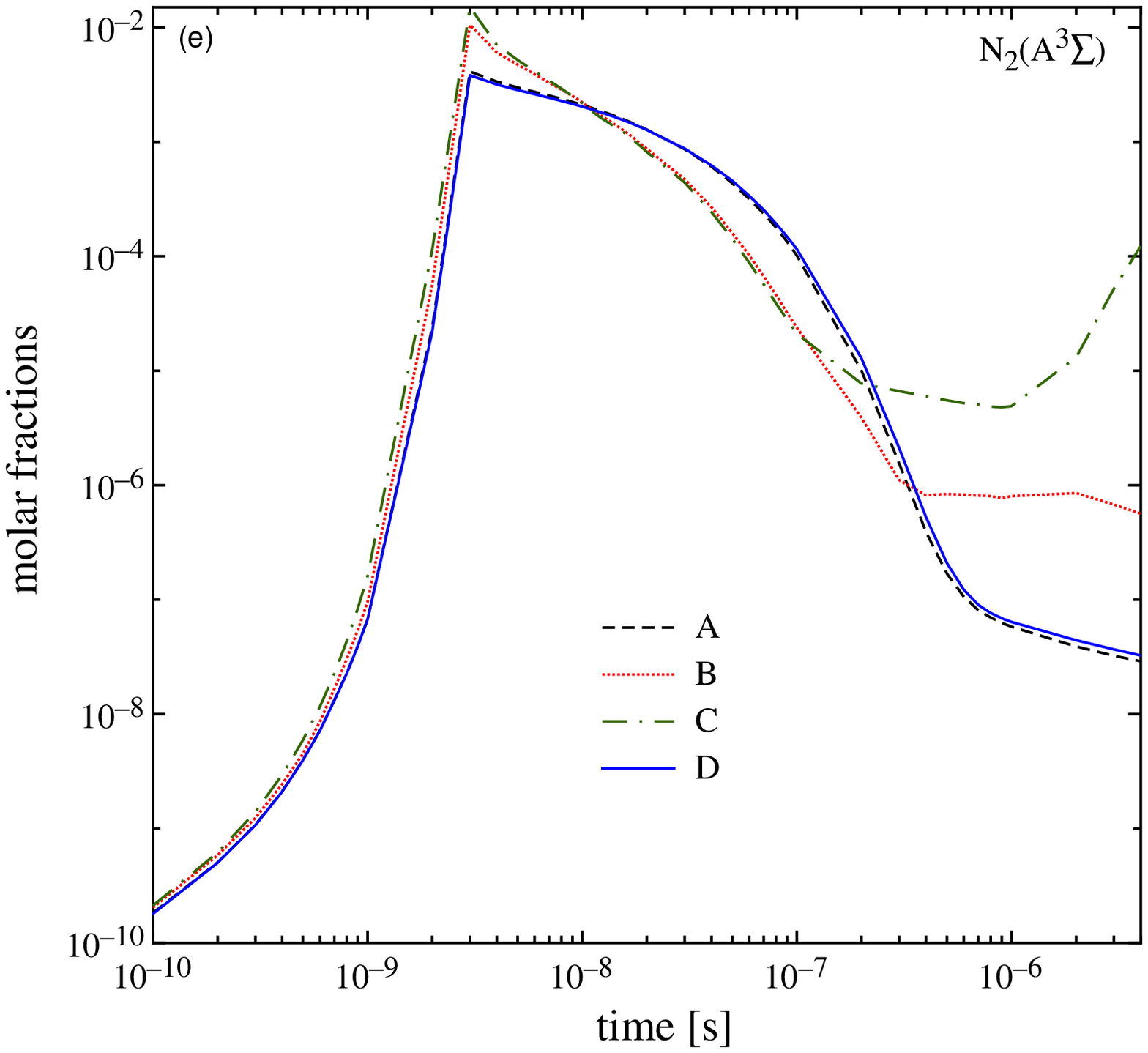} & \includegraphics[scale=.35,angle=0]{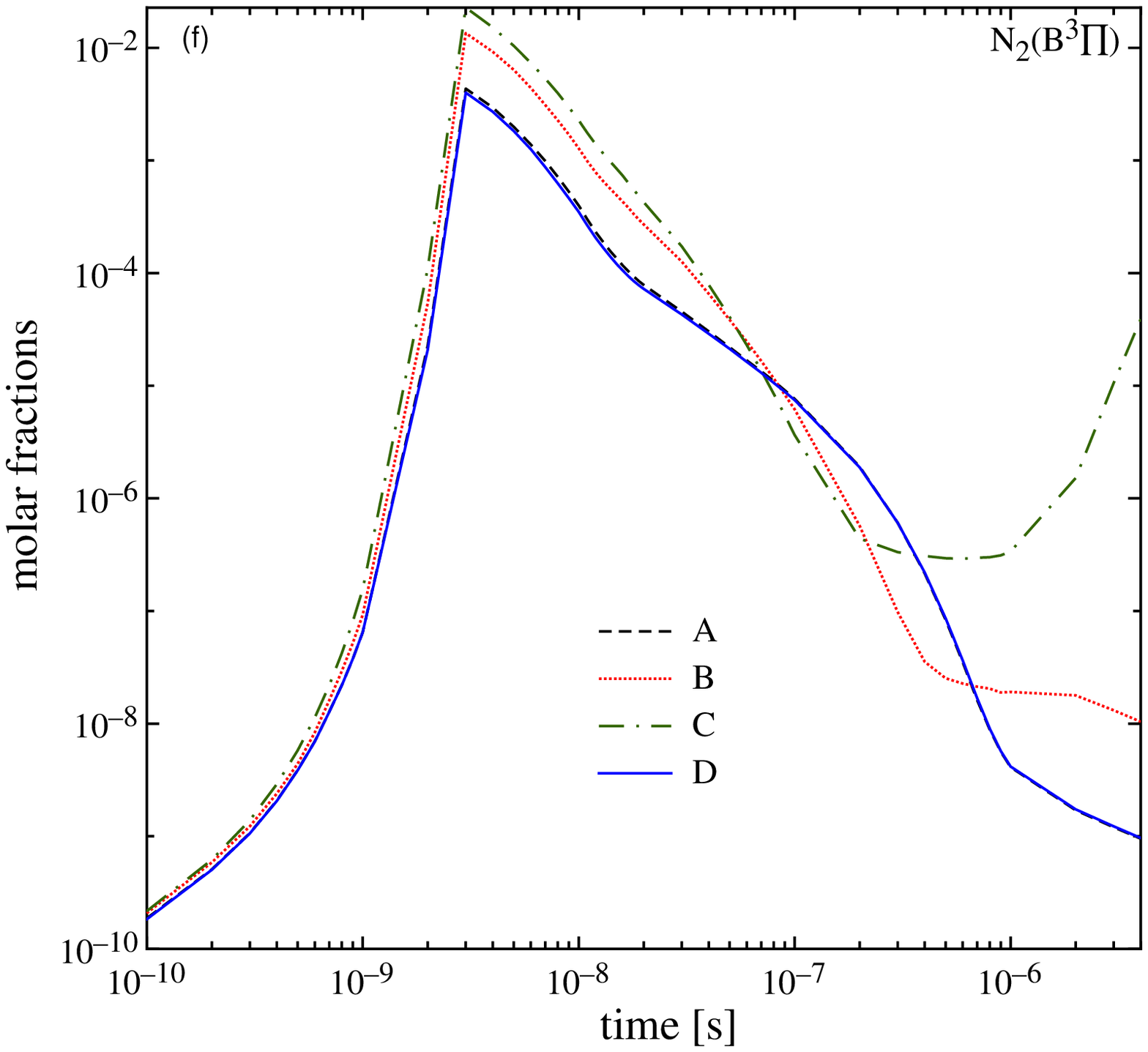} \\
\end{tabular}
\end{indented}
\caption{Time evolution of (a) mean electron energy; (b) nitrogen vibrational temperature; (c) free electrons molar fraction; (d) nitrogen atoms molar fraction; (e) and (f) molar fraction of N$_2$ triplet states for the four models considered. \label{fig:el_en_time}}
\end{figure}

The eedf starts to be affected by the presence of vibrationally excited states, as well as electronically excited states, when their population starts to be important. As an example, the vibrational energy supplied by electron impact is proportional to the electron density times the e-V rate; therefore, by considering the maximum value obtained for the electron density ($n_e \approx 1.5\times10^{16}$~cm$^{-3}$), the characteristic time for vibrational excitation is given by:
\begin{equation}
t_{qsv} = (n_e K_{eV})^{-1} \approx 3.3\ \textrm{ns} \,.\label{eq:tqsv}
\end{equation}
We approximately recover these times in Fig.~\ref{fig:el_en_time}(a) where the average energy of electrons is shown as a function of time for models A to D.

\begin{figure}
\begin{tabular}{ccc}
\includegraphics[scale=.3,angle=0]{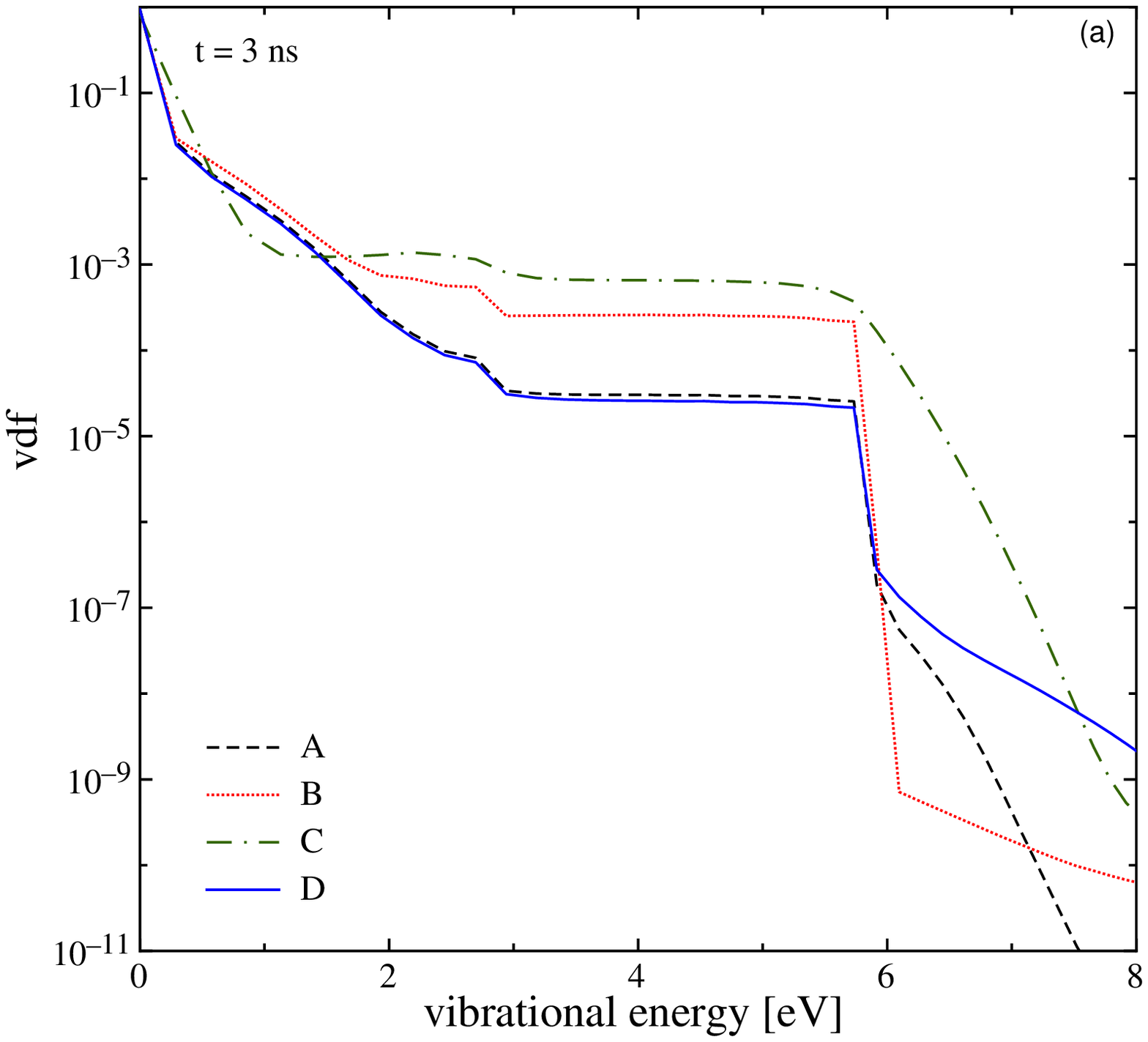} & \includegraphics[scale=.3,angle=0]{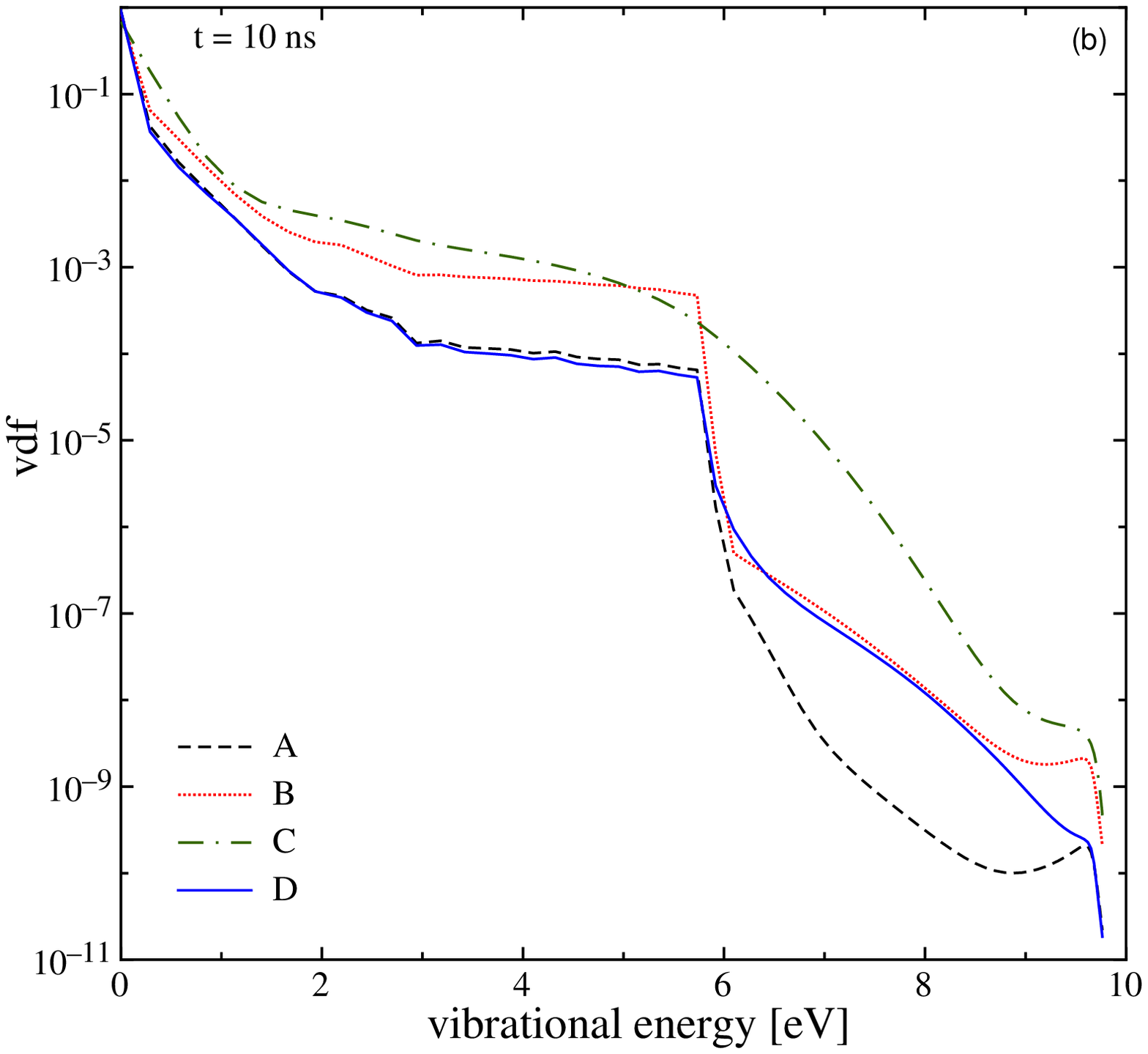} & \includegraphics[scale=.3,angle=0]{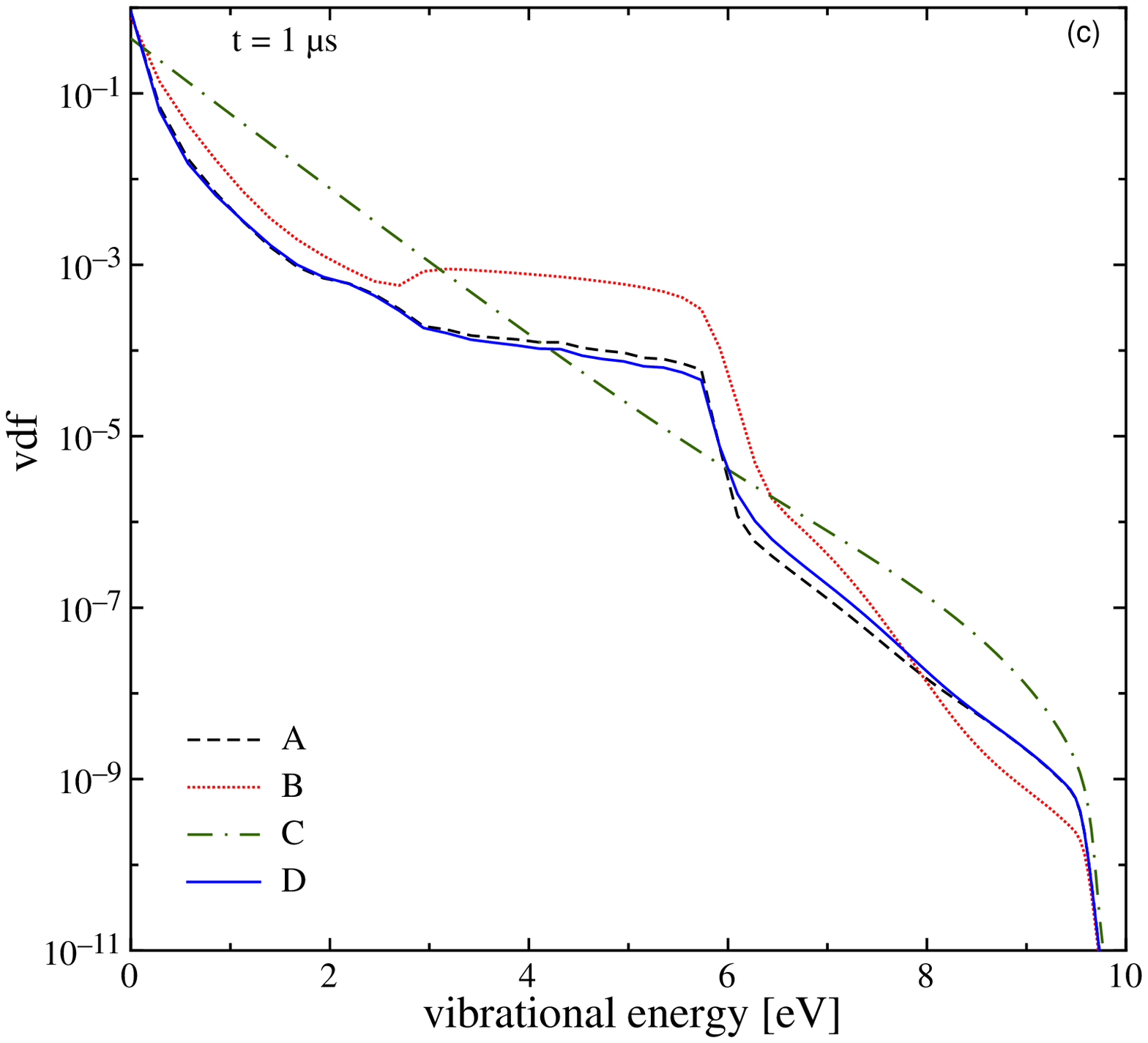}\\
\end{tabular}
\caption{Vibrational distributions function of N$_2$ molecules at three characteristic times and for the four models considered in the text. \label{fig:vdf_N2}}
\end{figure}

We start by discussing macroscopic quantities which can be useful for understanding the behaviour of the vdf and eedf under discharge and post-discharge conditions. Figure \ref{fig:el_en_time}(a), as already noted, reports the time evolution of the electron average energy calculated from the eedf as a function of the different models. We can see that in the range 0.1-1 ns the mean electron energy remains constant which is typical of an eedf in the cold gas approximation, \emph{i.e.} without the action of superelastic collisions, while the energy undergoes a sudden increase in the time interval 1-3 ns due to the excitation of vibrational and electronic states. For $t = 3$~ns, the electron mean energy abruptly decreases as the applied field is switched off. For $t > 3$~ns, the mean electron energy is sustained by the vibrational and electronic energy content of molecules, which through superelastic collisions, heat cold electrons. Our results depend strongly on the model used for the electron-molecule cross sections, models A and D predict values of the average energies lower than the corresponding ones from models B and C. The differences in the average energies for the four models can be recovered from the corresponding differences in the vibrational temperature calculated from the first two vibrational levels (see Fig.~\ref{fig:el_en_time}(b)). We note that case C deviates from the results of cases A, B and D mainly due to the poor representation of the vibrational excitation with the scaling-law given by Eq. (\ref{eq:scalinglaw3}) which generates much pumped vibrational distribution functions (see Fig.~\ref{fig:vdf_N2}). These differences are important under post-discharge condition but remain negligible during the discharge. Inspection of the results shows similarities between the mean electron energy obtained from the four models. The vibrational temperature initially shows a quasi-stationary value, then grows for nearly 3 ns, where the mean electron energy shows a peak. In the post-discharge, the vibrational temperature still increases, unlike the mean electron energy, as it is sustained by chemical processes (recombination) and electronically excited states. \textcolor{red}{Note that the temporal trend of 1-0 vibrational temperature is practically the same as the energy stored in the vibrational mode. In the long time range ($10^{-4}-10^{-3}$ s) model B and C show a decrease of the vibrational energy, while for models A and D the cooling phase is moved to longer times. At longer times the vibrational temperature will converge to the gas temperature.}

\textcolor{green}{Fig.~\ref{fig:el_en_time}(c) and (d) report the time evolution of free electron and atomic nitrogen molar fractions.} Again models A and D give very similar results while models B and C predict a much higher concentrations of electrons and nitrogen atoms due to the corresponding differences in eedf as well as in the the electron density. In particular the electron density is affected by electron impact ionization collisions involving the plateau of the vibrational distribution functions which are much more pumped in the cases B and C (see Fig. \ref{fig:vdf_N2}). The lower recombination rates of cases B and C compared with the cases A and D are probably due to additional ionization channels promoted by the plateau of the vibrational distribution function as well by electron impact collisions with electronically excited states. The behaviour of the two triplets of nitrogen molecules shown in Fig.~\ref{fig:el_en_time}(e) and (f) is also interesting: their concentration follows the same trend as the electron density and reaches a maximum value at the end of the ns discharge. Large differences are observed for $t > 1\ \mu$s where models B and C predict much higher concentrations due to more effective recombination. The behaviour of the two triplets of nitrogen molecules  is also interesting: their concentration follows the same trend as the electron density under discharge conditions reaching a maximum value at the end of the ns pulse. Large differences are observed for $t > 1$ $\mu$s where models B and C predict much higher concentrations as a consequence of 3-body N atom recombination forming N$_2(B)$ (and then N$_2(A)$ through radiative decay). Therefore N$_2(A)$ and N$_2(B)$ states follow the behaviour of nitrogen atoms as reported in Fig.~\ref{fig:el_en_time}(d).

Figure \ref{fig:vdf_N2} shows the behaviour of the vdf obtained by the four models at different characteristic times: (a) at the end of the discharge pulse; (b) at $t = 10$~ns; and (c) at $t = 1\ \mu$s in the post-discharge. Models A and D give similar vdfs up to 6 eV, for $t = 3$ and 10 ns, but predict large differences above 6 eV. This behaviour reflects the limitations of the scaling-law of Eq.~(\ref{eq:scalinglaw}), as can be observed in Fig. \ref{fig:scalinglaw}. Models B and C behave similarly up to 10 ns. At 1$\  \mu$s, model C predicts a quasi-Boltzmann distribution but model B shows a well-developed plateau, in line with models A and D, but at much higher $v$ than in these cases.

\begin{figure}
\begin{indented}
\item[]
\begin{tabular}{cc}
\includegraphics[scale=.35,angle=0]{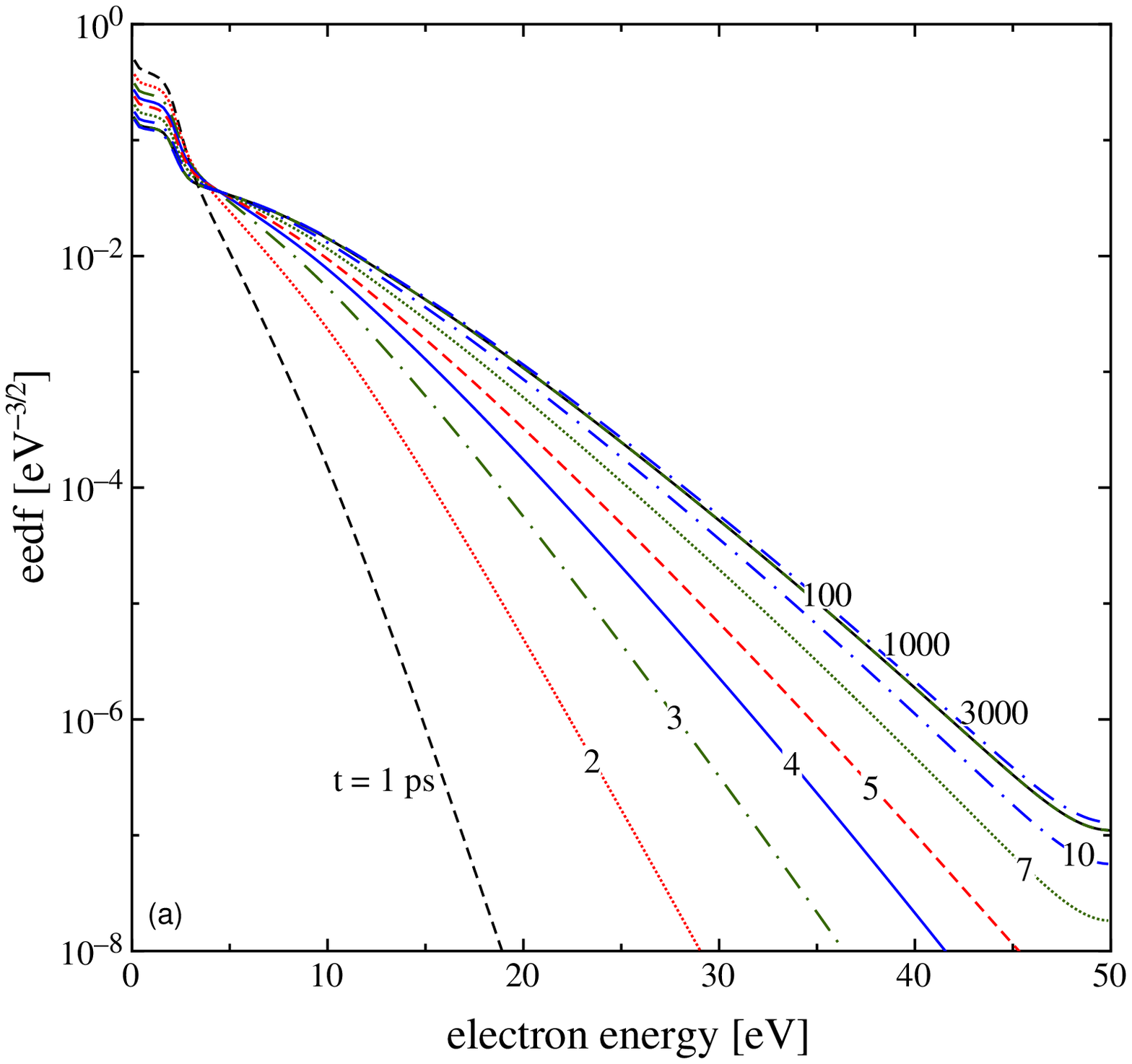} & \includegraphics[scale=.35,angle=0]{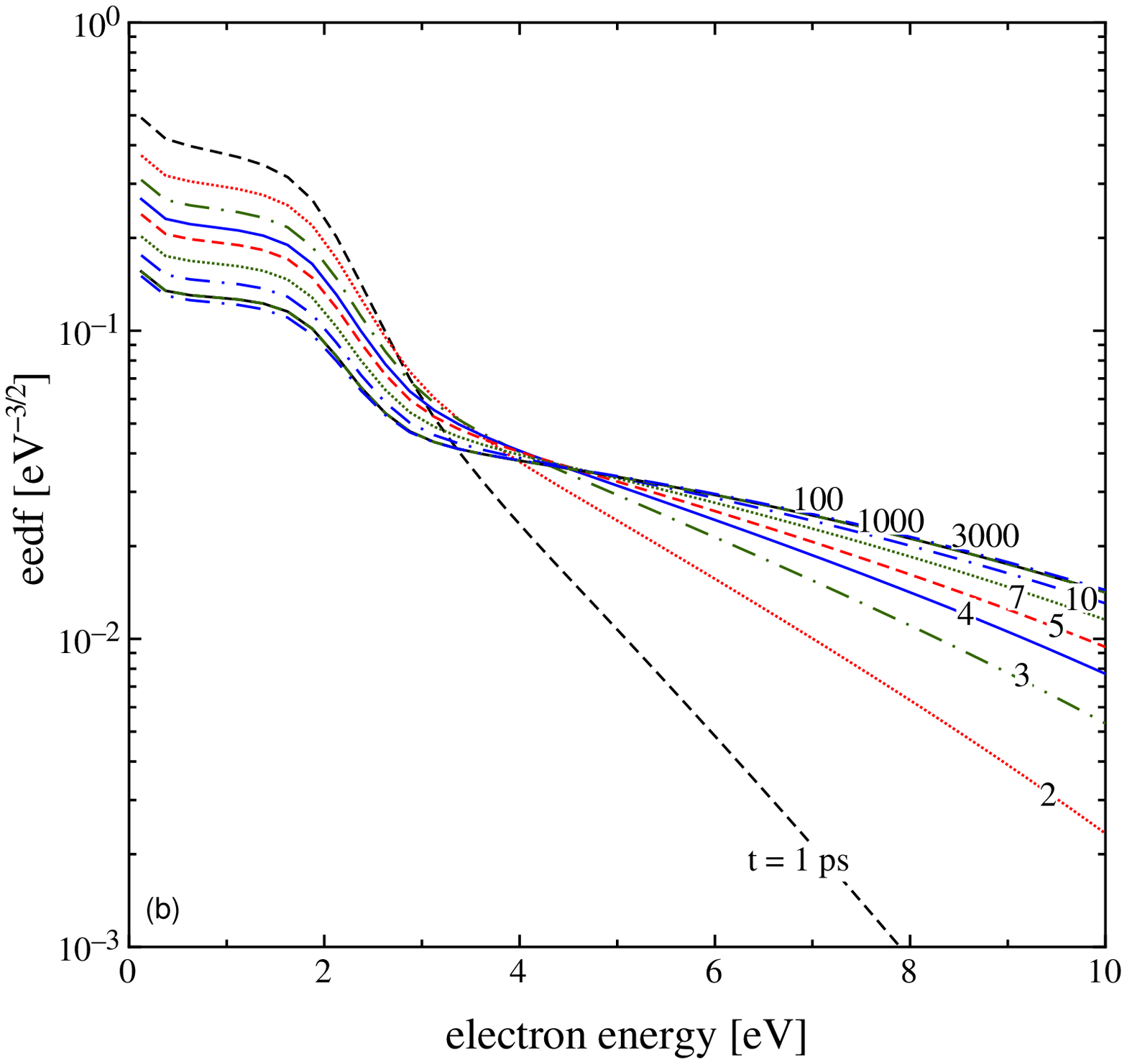}\\
\end{tabular}
\end{indented}
\caption{(a) Time evolution of eedf under discharge conditions according to model D. (b) The same plot in (a) in the energy range 0-10 eV. \label{fig:eedf}}
\end{figure}

\begin{figure}
\begin{indented}
\item[]
\begin{tabular}{cc}
\includegraphics[scale=.35,angle=0]{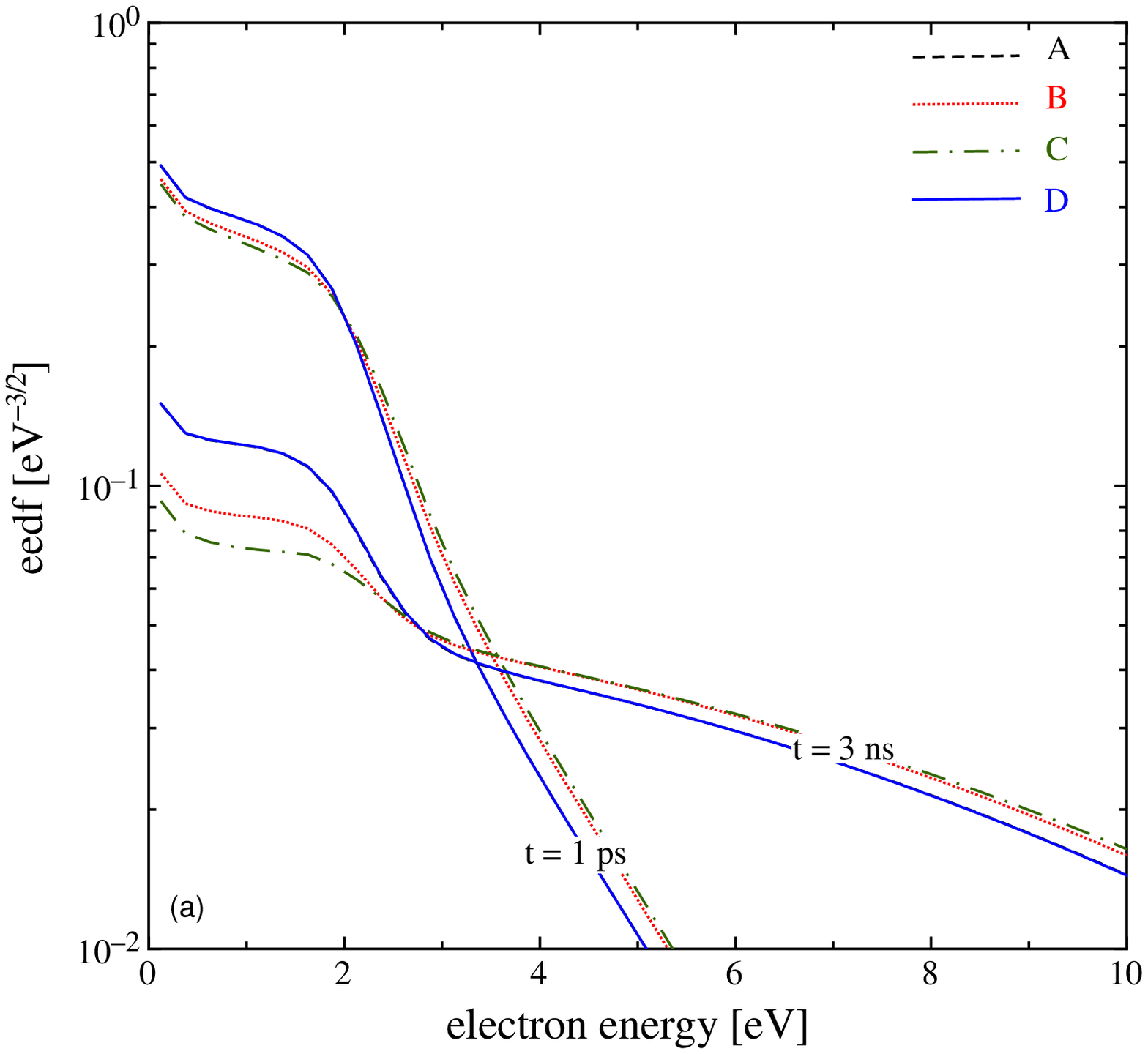} & \includegraphics[scale=.35,angle=0]{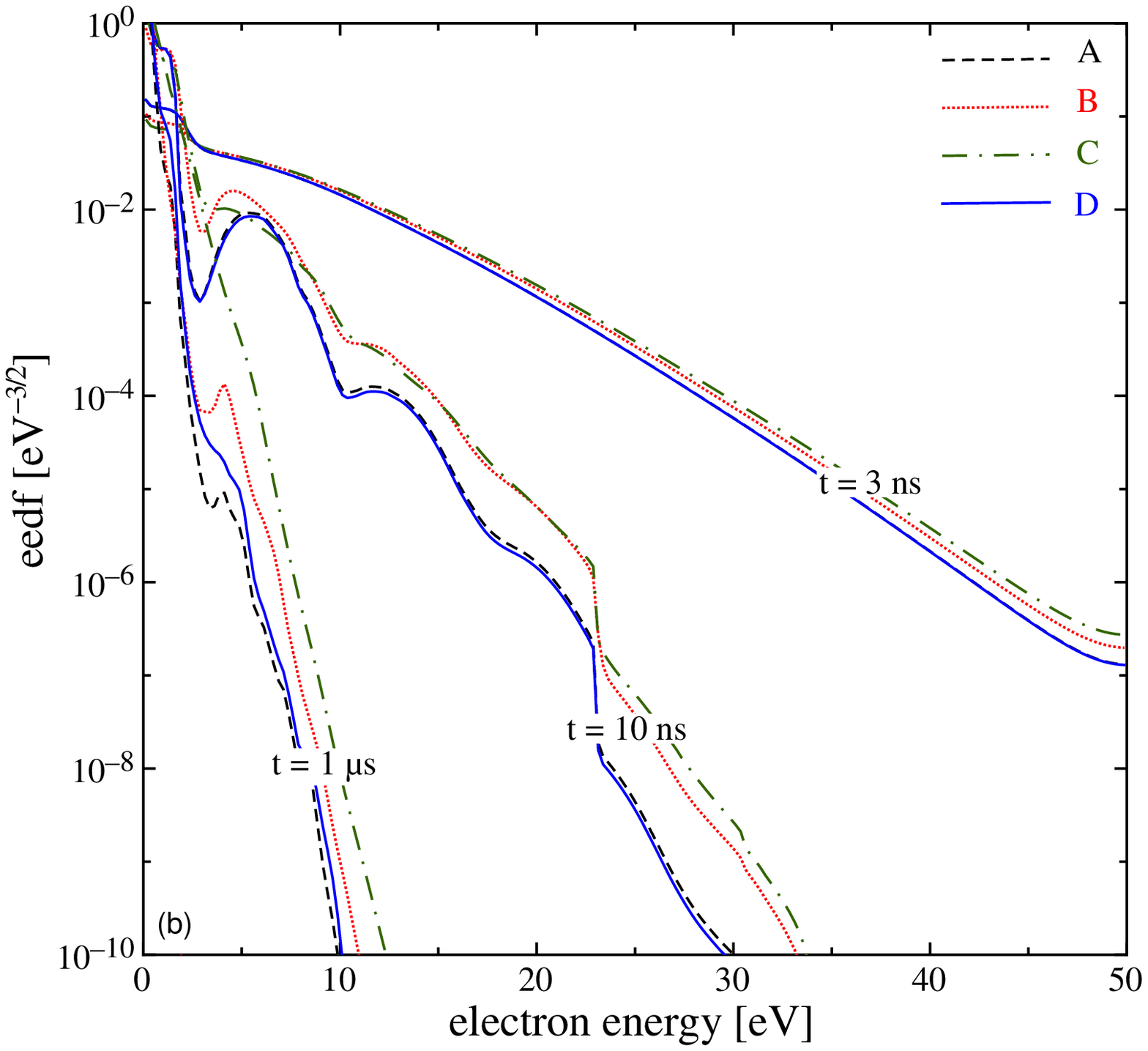}\\
\end{tabular}
\end{indented}
\caption{(a) eedf at two times during the discharge; (b) full range eedf during the post-discharge for the four models. \label{fig:eedf_bis}}
\end{figure}

Let us now consider the time evolution of eedf under discharge and post discharge conditions. Figure \ref{fig:eedf} illustrates the situation during the discharge. The evolution of the eedf towards a quasi-stationary state is rather regular. After 0.1 ns, the eedf reaches a quasi-stationary value and, in this regime, superelastic collisions do not affect the distribution due to the high value of the electron energy. As shown in Fig. \ref{fig:eedf_bis}, the four models give small differences in the eedf, especially in the low-energy region, being dominated by e-V inelastic processes. The situation completely changes in the post-discharge regime where the decrease of average energy enhances the role of superelastic collisions from electronic excited states, \emph{i.e.}:
\begin{eqnarray}
e(\epsilon) + \textrm{N}_2(\textrm{A}\ ^3\Sigma) &\to& e(\epsilon+\epsilon^*) + \textrm{N}_2(\textrm{X}\ ^1\Sigma^+_g, v)\,, \\
e(\epsilon) + \textrm{N}_2(\textrm{B}\ ^3\Pi)         &\to& e(\epsilon+\epsilon^*) + \textrm{N}_2(\textrm{X}\ ^1\Sigma^+_g, v)\,, \\
e(\epsilon) + \textrm{N}_2(\textrm{C}\ ^3\Pi)        &\to& e(\epsilon+\epsilon^*) + \textrm{N}_2(\textrm{X}\ ^1\Sigma^+_g, v)\,,
\end{eqnarray}
where $\epsilon^*$ is the threshold energy of the corresponding transition, that are, relative to the ground state respectively 6.17 eV, 7.35 eV and 11.03 eV. At $t=10$~ns these processes are responsible of the peak at $\epsilon\approx6$ eV, while the second peak at $\epsilon\approx12$ eV is due to electrons in the peak at $\epsilon\approx6$ eV subjected to another superelastic collision. It should be noted that the peaks are quite large due to the superposition of the different superelastic collisions. The effect of transition from the C$\ ^3\Pi$ state are visible as small shoulders around 10 eV and 20 eV. These maxima are strongly smoothed by the model C due to the higher mean electron energy, which hides the role of superelastic collisions. Similar results are observed for $t = 1\ \mu$s, where only the contribution of the A$^3\Sigma$ state is clearly evident.

These results clearly demonstrate that the use of complete sets of cross sections, as well as the coupling between vibrationally and electronically excited states, is very important. This is especially true in the post-discharge phase. Neglecting these aspects will lead to non-negligible errors in any model. The results in Fig.~\ref{fig:eedf_bis} warns on the use of the Maxwell distribution function for calculating electron-molecule rate coefficients. The results reported in Fig.~\ref{fig:scalinglaw}, as well as in Ref.~\cite{0963-0252-21-5-055018} can be used only for particular conditions \emph{i.e.} at very high ionization degree when electron-electron Coulomb collisions dominates.

\section{Conclusions and Perspectives \label{sec:conclusions}}

We have presented the dependence of the macroscopic (mean electron energy, vibrational temperature) and microscopic (vdf, eedf) quantities in a nitrogen discharge plasma on different models for the e-V transitions. As a benchmark model we use the recent database calculated by Laporta \emph{et al.}. Their cross sections are compared with the corresponding ones obtained using the analytic expression of Bourdon \emph{et al.} that completes the data-set by scaling-laws the ground state cross sections, and other scaling-laws, which restricts the number of interconnecting vibrational levels. Use of these restrictive models give results far from the benchmark model while the scaling-law of Bourdon \emph{et al.} gives good agreement, particularly for mean electron energies larger than 0.5 eV. Note that, for sake of coherence, all the scaling-laws have been applied to the data set of Laporta \emph{et al}. It should be pointed out that the scaling-law of Bourdon \emph{et al.} has been derived from the accurate set of e-V cross sections of Huo \emph{et al.}. This implies that in general complete, \emph{ab initio} e-V cross sections should be used in preference to scaled ones if one wants a good representation of the vibrational kinetics of diatomic molecules. In this context we note that complete cross e-V cross section sets have recently prepared for CO~\cite{0963-0252-21-4-045005} and O$_2$~\cite{0963-0252-22-2-025001}. In future we plan to use these cross sections to extend our approach to the study of more complex mixtures such as air and air/hydrocarbon involved in combustion processes, where the presence of excited states is very important.

In these new studies we plan to apply a more realistic form of the applied electric field by substituting the present Heaviside-tipe temporal evolution of $E/N$ with  shapes derived from the experiments~\cite{0022-3727-39-16-R01, 0022-3727-47-11-115201, 0741-3335-57-1-014001}. The new results should confirm the differences obtained in the present work in the comparison of models A-C (\emph{i.e.} on the choice of e-V scaling-laws) even though we expect  no-negligible variations in the absolute values of the relevant macroscopic and microscopic quantities following the selected different experimental time dependent voltage forms.

\ack
This work received funding from the European Union's Seventh Framework Programme (FP7/ 2007-2013) under Grant agreement no. 242311 and from Regione Puglia under project INNOVHEAD-Avviso Miur n713/Ric.

\section*{References}
\addcontentsline{toc}{section}{References}

\bibliographystyle{is-unsrt}{}  


\end{document}